\documentclass[12pt,preprint]{aastex}

\def\lesssim{\mathrel{\hbox{\rlap{\hbox{\lower4pt\hbox{$\sim$}}}\hbox{$<$}}}}
\def\gtrsim{\mathrel{\hbox{\rlap{\hbox{\lower4pt\hbox{$\sim$}}}\hbox{$>$}}}}

\begin{document}
\title{Very Early Optical Afterglows of Gamma-Ray Bursts: Evidence for 
Relative Paucity of Detection}
\author{Peter W.A. Roming\altaffilmark{1}, Patricia Schady\altaffilmark{1,2}, Derek
B. Fox\altaffilmark{3,1}, Bing Zhang\altaffilmark{4}, Enwei Liang\altaffilmark{4}, Keith O. Mason\altaffilmark{2},
Evert Rol\altaffilmark{5}, David N. Burrows\altaffilmark{1}, Alex J. Blustin\altaffilmark{2}, Patricia
T. Boyd\altaffilmark{6,7}, Peter Brown\altaffilmark{1}, Stephen T. Holland\altaffilmark{6,8},
Katherine McGowan\altaffilmark{2}, Wayne B. Landsman\altaffilmark{6}, Kim L.  Page\altaffilmark{5}, James
E. Rhoads\altaffilmark{9}, Simon R. Rosen\altaffilmark{2}, Daniel Vanden Berk\altaffilmark{1}, 
Scott D. Barthelmy\altaffilmark{6}, Alice A.
Breeveld\altaffilmark{2}, Antonino Cucchiara\altaffilmark{1}, Massimiliano De Pasquale\altaffilmark{2},
Edward E.  Fenimore\altaffilmark{10}, Neil Gehrels\altaffilmark{6}, Caryl Gronwall\altaffilmark{1}, Dirk
Grupe\altaffilmark{1}, Michael R. Goad\altaffilmark{5}, Mariya Ivanushkina\altaffilmark{1,11}, Cynthia
James\altaffilmark{2}, Jamie A. Kennea\altaffilmark{1}, Shiho Kobayashi\altaffilmark{1}, Vanessa
Mangano\altaffilmark{12}, Peter M\'esz\'aros\altaffilmark{1}, Adam N. Morgan\altaffilmark{1}, John
A. Nousek\altaffilmark{1}, Julian P. Osborne\altaffilmark{5}, David M. Palmer\altaffilmark{10}, Tracey
Poole\altaffilmark{2}, Martin D. Still\altaffilmark{6,8}, Gianpiero Tagliaferri\altaffilmark{13},
Silvia Zane\altaffilmark{2}}

\altaffiltext{1}{ Department of Astronomy \& Astrophysics,
Pennsylvania State University, 525 Davey Lab, University Park, PA
16802, USA; Corresponding author's e-mail: roming@astro.psu.edu}
\altaffiltext{2}{ Mullard Space Science Laboratory, University College London,
Holmbury St. Mary, Dorking, Surrey RH5 6NT, UK} 
\altaffiltext{3}{ Division of Physics, Mathematics and Astronomy, 105-24,
California Institute of Technology, Pasadena, CA 91125, USA} 
\altaffiltext{4}{ Department of Physics, University of Nevada, Las Vegas, 4505
Maryland Parkway, Las Vegas, NV 89154-4002, USA} 
\altaffiltext{5}{ Department of Physics and Astronomy, University
of Leicester, University Road, Leicester LE1 7RH, UK } 
\altaffiltext{6}{ NASA/Goddard Space Flight Center, Greenbelt, MD 20771, USA} 
\altaffiltext{7}{ Joint Center for Astrophysics, University of Maryland, 1000
Hilltop Circle, Baltimore, MD 21250, USA} 
\altaffiltext{8}{ Universities Space Research Association, 10227 Wincopin Circle,
Suite 212, Columbia, MD 21044, USA} 
\altaffiltext{9}{ Space Telescope Science Institute, 3700 San Martin Drive,
Baltimore, MD 21218, USA} 
\altaffiltext{10}{ Los Alamos National Laboratory, P.O.
Box 1663, Los Alamos, NM 87545, USA} 
\altaffiltext{11}{ Department of Physics and
Astronomy, Brigham Young University, N283 ESC, Provo, UT 84602, USA}
\altaffiltext{12}{ INAF-Instituto di Astrofisica Spaziale e Fisica Cosmica 
Sezione di Palermo, via Ugo La Malfa 153, I-90146 Palermo, Italy}
\altaffiltext{13}{ INAF-Osservatorio Astronomico di Brera, Via Bianchi 46, 23807
Merate, Italy}

\begin{abstract}
Very early observations with the Swift satellite of $\gamma$-ray burst 
(GRB) afterglows reveal that the optical component is not detected in
a large number of cases. This is in contrast to the bright optical 
flashes previously discovered in some GRBs (e.g. GRB 990123 and GRB 
021211). Comparisons of the X-ray
afterglow flux to the optical afterglow flux and prompt $\gamma$-ray
fluence is used to quantify the seemingly deficient optical, 
and in some cases X-ray, light at these early epochs. This 
comparison reveals that 
some of these bursts appear to have higher than normal $\gamma$-ray 
efficiencies. We discuss possible mechanisms and their feasibility 
for explaining the apparent lack of early optical emission. The 
mechanisms considered include: foreground extinction, circumburst 
absorption, Ly-$\alpha$ blanketing and absorption due to high redshift, 
low density environments, rapid temporal decay, and intrinsic weakness 
of the reverse shock. Of these, foreground extinction, circumburst 
absorption, and high redshift provide the best explanations for most of 
the non-detections in our sample. There is tentative evidence of 
suppression of the strong reverse shock emission. This could be because of 
a Poynting-flux-dominated flow or a pure non-relativistic hydrodynamical 
reverse shock.

\end{abstract}

\keywords{gamma-rays: bursts}

\section {Introduction}
The afterglow discoveries of 1997 revealed that gamma-ray bursts
(GRBs) are the brightest explosions in the universe. They occur at
cosmological distances (Metzger et al. 1997) and produce long-lived
emission across the electromagnetic spectrum, from the X-ray band
(e.g. Costa et al. 1997) through optical (e.g. Van Paradijs et al. 1997) to
radio (e.g. Frail et al.  1997) wavelengths. While X-ray afterglows of GRBs
are detected in nearly all cases (De Pasquale et al. 2003 - hereafter
D03; Burrows et al. 2006), their optical afterglows often remain 
undetected even in deep searches (Roming \& Mason 2006 - hereafter RM06). 
Identifying and characterizing these ``dark" bursts has been
difficult due to the delays (typically hours) occurring between the
detection of the GRB and the execution of the first ground-based
observations. 

Classical explanations for dark bursts include: absorption at the burst 
location (Piro et al. 2002; Lazzati, Covino, \& Ghisellini 2002; 
Djorgovski et al. 2001; Fynbo et al. 2001 - hereafter F01), low density
environment (Frail et al. 1999; Groot et al. 1998a; Taylor et
al. 2000), rapid temporal decay (Groot et al. 1998a), foreground extinction
(Taylor et al. 1998), Ly-$\alpha$ blanketing and absorption due
to high redshift (F01; Groot 1998b), or intrinsic faintness (Lazzati, 
Covino, \& Ghisellini 2002;
F01). Previous authors have quantified the degree of optical darkness
expected based on the upper limit on the afterglow flux (ULAF; Rol et
al. 2005a) or the optical-to-X-ray spectral index ($\beta_{OX}$;
Jakobsson et al. 2004 - hereafter J04) and show that most ``dark"
bursts can be explained by adverse observing conditions. 

Observations of GRBs
with NASA's Swift satellite (Gehrels et al. 2004) are providing prompt
few-arcminute $\gamma$-ray localizations, rapid few-arcsecond X-ray
positions, and rapid and extensive follow-up in the X-ray, UV,
optical, and radio bands (e.g. Gehrels et al. 2005; Cameron \& Frail
2005).  Thirteen of these bursts include extraordinary optical upper
limits at very early epochs after the burst. This is in contrast to
the bright optical flashes discovered to accompany some GRBs
(e.g. Akerlof et al. 1999; F01; Fox et al. 2003; Li et
al. 2003) in the pre-Swift era. 

In this paper we report the very early observations
of GRB afterglows
by the Swift UV/Optical Telescope (UVOT; Roming et al. 2005a) and the 
associated
very tight early optical upper limits. 
In Section 2, we present the observations and data reduction methods. 
In Section 3, we combine the Swift optical, X-ray, and $\gamma$-ray data 
sets to quantify the apparent lack of optical, and in few instances X-ray, 
flux at these early epochs. 
We 
discuss
possible mechanisms
for explaining the apparent lack of early optical afterglows.
In Section 4, we provide our conclusions.

\section{Observations \& Reductions}
Between 24 January and 30 June, 2005, the Swift Burst Alert Telescope
(BAT; Barthelmy et al. 2005) detected and localized 26 GRBs which were
thereafter observed by the Swift X-Ray Telescope (XRT; Burrows et
al. 2005a) and UVOT.
Identification of the fading X-ray afterglow in each of these
cases has allowed the distribution of positions accurate to better
than 6 arcseconds within minutes to hours of the burst. The positions
of 13 of these afterglows were observed with the UVOT less than one
hour after the burst, without optical counterparts being
found (see Table 1 \& Figure 1). Seven of the thirteen afterglows were
observed by Swift within two minutes of the burst onset with no
optical emission detected by UVOT. For comparison purposes, we have
included six bursts that were detected by the UVOT (see Table
1). 

Table 1 provides the basic properties of the 19 bursts in the
Swift UVOT sample. The XRT data were processed using version 2.0 of the 
Swift software.
Cleaned event lists were read into XSELECT, where source and background
spectra and light-curves were extracted. If the data appeared piled-up in
Photon Counting (PC) mode (i.e. if the count rate was $\gtrsim 0.6
{\rm count} \ {\rm s}^{-1}$), 
an annular extraction region was used, with the inner radius dependent on
the level of pile-up; otherwise, a circle of radius $\sim$30 pixels was chosen
(1 pixel = 2.36 arcseconds). The background files were obtained from nearby,
`source-free' areas of sky. The X-ray light-curves were modelled with a 
combination of single and
broken power-laws and the models then used to estimate the count rates at
1 and 11 hours. The spectra were fitted in XSPEC v11.3 and used to
obtain a mean count-rate to unabsorbed flux conversion.
The gamma-ray fluence was calculated in the 15-350 keV energy band using 
the best-fit spectral parameters. For GRBs observed with other telescopes, 
a break energy at 250 keV with the low and high photon energy indices set 
to -1 and -2.3, respectively (Band et al. 1993) was assumed.
The 3$\sigma$ upper limit for each burst was ascertained
using a 3 and 6 arcsecond diameter aperture (for the optical and UV
filters, respectively) around the XRT position and then correcting for
galactic extinction. All upper limits and magnitudes were determined
using the UVOT V-filter finding chart, which occurs after the
spacecraft has settled on the target.  R-band values were derived from
the UVOT V-bandpass by extrapolating to the effective R-band
wavelength with the assumption of a $\nu^{-1}$ spectral shape. Errors
on the R-band magnitudes range from $\pm 0.22$ to $\pm 0.53$. The zero 
points from Fukugita, Shimasaku, \& Ichikawa (1995) were 
used to convert the magnitude to 
fluxes. The bluest magnitude for each burst, established by ground based
observers, was obtained from the GRB Coordinate Network (GCN)
Circulars (Barthelmy et al. 1995; Barthelmy et al. 1998). Redshifts
were obtained from both the GCN and UVOT observations.

\section{Discussion}
Beppo-SAX and HETE-2 satellite studies (J04) led us to expect that the
UVOT would readily detect the prompt optical emission of most GRBs and 
their afterglows\footnote{We point out that Kehoe et al. 2001 showed that 
not all GRBs have GRB 990123-like optical flashes}. However, the upper 
limits provided here show the
UVOT bursts are not detected at or blue-wards of the V-band at these early
times. The speed of response and depth of UVOT observations are
superior to the observations of all but a few previous GRBs. Moreover,
accurate XRT positions make us confident that no bright optical
afterglows were masked due to confusion with bright, nearby stars. The
sensitivity and multiwavelength nature of our observations thus offers
a unique opportunity to study the optical deficiency of these early
afterglows.

As an initial test of the optical deficiency of our sample we employed the
quantified ULAF and $\beta_{OX}$ methods. The $\beta_{OX}$ method is a rapid 
technique for ascertaining the 
potential
darkness of Swift bursts. Those bursts with 
$\beta_{OX} < 0.50$ are classified as dark, while those bursts in the range 
of $0.50 < \beta_{OX} < 0.55$ are classified as potentially dark. The ULAF 
method is a more thorough approach to dark burst classification. It utilizes 
the temporal ($\alpha$) and spectral ($\beta$) indices for determining the 
minimum and maximum value for the electron index ($p$). Eight cases of the 
standard afterglow model are tested and the resultant maximum and minimum X-ray 
flux are extrapolated to the optical epoch. Optical flux that falls below the 
minimum extrapolated X-ray flux is considered dark.
Using the ULAF method, we 
found that only GRB 050219B could be classified as dark (a determination of 
the darkness of GRB 050421 could not be made with the ULAF method). The
$\beta_{OX}$ method revealed that GRBs 050315, 050319, 050326, 050412, and
050505 were classified as dark (RM06; see Figure 2). Even though the ULAF or 
$\beta_{OX}$ classification schemes do not categorize most bursts in our 
sample as dark, it does not explain why the early optical 
afterglows are 
apparently lacking.
For both the $\beta_{OX}$ and ULAF methods, it
is assumed that GRBs follow the standard fireball model (RM06). 
Recent Swift observations show much more complicated features in 
the early afterglow phase
(Nousek et al. 2006 - hereafter N06; Zhang et al. 2006a - hereafter Z06a), 
which calls for a 
more detailed study of the optical deficiency in the early phases.

To quantify the apparent
lack of optical and X-ray light at these
early epochs in the Swift bursts, we combine broadband UVOT
observations with BAT and XRT data and compare the burst's X-ray
afterglow flux at one hour after the burst ($F_{x,1}$) to its prompt
$\gamma$-ray fluence ($S_\gamma$) and the optical afterglow flux at
one hour ($f_{R,1}$).  This is done because the ratio
$S_{\gamma}/S_{x,1}$ (where $S_{x,1}=F_{x,1}~(1 {\rm hr})$ is the
1-hour X-ray fluence) measures the $\gamma$-ray efficiency of the 
event; while the ratio, $F_{x,1}/f_{R,1}$, measures the amount of
optical flux constrained by selective extinction (J04;
D03). $S_{\gamma}$ is the natural measure of the radiated energy of
the GRB. The X-ray band flux at a certain epoch, e.g. $F_{x,1}$ (or
fluence $S_{x,1}$) is a good indication of the fireball's 
blastwave kinetic energy, especially if the cooling frequency is below
the band (Kumar 2000; Freedman \& Waxman 2001; Berger, Kulkarni, \&
Frail 2003; Lloyd-Ronning \& Zhang 2005). Although the fireball bulk
Lorentz factor is different 
during the prompt gamma-ray phase and one hour after the burst
trigger, the energy per solid angle along the viewing direction is
essentially independent of the value of the bulk Lorentz factor and
the jet structure (Zhang \& M\'esz\'aros 2002). The physical jet angle
also does not change between the epoch of the prompt phase (when
$S_\gamma$ is measured) and 1-hour after the trigger (when $F_{x,1}$
and $S_{x,1}$ is measured). Finally, the value $S_\gamma/S_{x,1}$ only
weakly depends on the unknown redshift $z$ when the X-ray band is
above the cooling frequency (up to a correction factor of order
unity; see e.g. Zhang et al. 2006b - hereafter Z06b; Granot et al. 2006). 
As a result the ratio $S_\gamma/S_{x,1}$ (or $S_\gamma/F_{x,1}$), is 
an
indication of the gamma-ray efficiency. 

We choose 1-hour after the trigger as the fiducial epoch based on the
following considerations. First, we want to compare the Swift data
with the measurements of some pre-Swift bursts. The measurements of
pre-Swift bursts usually happened several hours after the burst
trigger. One therefore needs to extrapolate the late time flux to
early times (e.g. 1 hour). It is now clear that the early X-ray
afterglow lightcurves typically show a shallow decay phase extending
to about 1 hour (N06; Z06a). 
Extrapolating the pre-Swift data to an epoch much earlier than 1
hour inevitably leads to unreliable results, which would be
inappropriate for directly comparing against the Swift data
(cf. Fig.3 and Fig.4). On the other hand, the strength of Swift are the
early observations. In order to study the early afterglow properties it 
is unwise to extrapolate the data to later epochs. This is especially relevant
for upper limits since no decaying slope information is available. On 
the other hand, utilizing the Swift data at an epoch much earlier 
than 1 hour induces large uncertainties since the X-ray flux is dominated
by flares and/or a rapid decline phase which are thought to be associated
with the central engine rather than the afterglow. We
therefore find a compromise of both considerations to be the taking of 
1 hour as the
fiducial epoch. One caveat is that this epoch is typically the
transition time between the reverse shock component and the forward
shock component for some optical flashes. However, these optical
flashes are not commonly detected, especially in our sample. A
detailed GRB efficiency study at a much earlier and a much later epoch
has been presented elsewhere (Z06b).

In Figure 3, we can see that
most optically-detected GRBs prior to Swift fall into the gray band
--- which is twice the standard deviation (2$\sigma$) of
$S_{\gamma}/F_{x,1}$ for optically-detected GRBs --- thus 
seemingly
supporting
the idea that the efficiency of $\gamma$-ray production is roughly
constant. In a similar fashion, the broadband nature of synchrotron
emission 
implies
that the ratio of X-ray and optical emission at a
fixed epoch is roughly constant, except when selective extinction
suppresses the optical flux (J04; D03). Indeed, the X-ray and optical
fluxes of pre-Swift optically-detected GRBs are 
correlated (Fig. 4). 

Having quantified the apparent lack of optical and X-ray light, 
we now examine possible mechanisms. The first mechanism we consider is 
that of extinction by dust. For the case of foreground extinction, 
only galactic absorption in the optical can be considered since 
modeling intervening dust along the line of sight is problematic 
without an optical afterglow detection\footnote{A search for sources 
along the line of sight in the USNO-B catalogue revealed that there were 
no sources coincident with the position of the optically deficient GRBs in our sample}. 
The galactic absorption along the line of sight for most bursts in our
sample is relatively small. However, in the case of GRBs 050421, 050422, 
and 050509A the extinction is fairly large (see Table 1). In these cases 
it is not unreasonable to assume, at least in part,
that the afterglows of these bursts are 
extincted due to intervening dust. For the case of circumburst absorption, 
preliminary results by Schady et al. (2006a) on a small sample of 
optically detected Swift bursts suggests that this effect is likely 
small for most Swift bursts. This is in agreement with work done by 
Jakobsson et al. (2006) who suggest that only 20\% of Swift bursts are 
highly obscured. This value implies that 2-3 of the optically undetected burst in our 
sample should be significantly obscurred. An examination of Figure 4 
(right panel), which investigates the relationship between $F_{x}$ and 
$f_{R}$ at 11 hours (which is a reasonable time for assuming that the 
cooling frequency has passed through the optical band), reveals that 
GRBs 050219A, 050315, and 050412 are probably obscurred by circumburst
dust. We note that the optically detected burst GRB 
050319 is also below the 2$\sigma$-band.

The second mechanism we consider is that of Ly-$\alpha$ blanketing and 
absorption due to high redshift. The UVOT is capable of detecting bursts 
out to $z \sim 5$. As an example, GRB 060522 was recently detected by 
UVOT in the White-light filter at $z = 5.11$ (Holland 2006). The mean 
redshift of Swift localized bursts is $z_{mean} = 2.8$ with at least 
7\% of the bursts located at $z > 5$ (Jakobsson et al. 2006). Based on 
these criteria, and if we assume a fairly even distribution of bursts out 
to a redshift of $z \sim 5$ (see RM06), UVOT should detect most bursts 
in the first observations made with the White-light and V-filters out 
to $z \sim 4$. After this point the ability to detect an afterglow becomes
increasingly more difficult (see Figure 5). This equates to $\sim 30\%$, 
or 3-4 bursts in the sample, being optically undetected to UVOT due to Ly-alpha blanketing 
and absorption at high redshift. 

An additional mechanism for suppressing the optical afterglow is a rapid
temporal decay. The temporal decay profiles of a relatively small sample 
of detected Swift early optical afterglows are shallow ($\alpha \sim 1$)
in comparison to the canonical steep X-ray decay profiles ($\alpha \sim 
3$; Tagliaferri et al. 2005; N06; O'Brien et al. 2006; Z06a; 
RM06; Yost et al. 
2006). These shallow decays indicate that if detected bursts and optically 
deficient bursts behave similarly, then 
rapid temporal decay is not a significant consideration for our sample. 

Another possibility to account for a low afterglow flux level is a low 
medium density, if the X-ray band is below the cooling frequency. In fact 
about 1/4 of Swift X-ray afterglows are in this regime (Z06b). However, 
a low density medium does not give a straightforward 
interpretation of the suppression of an optical flash, if the latter is
attributed to a reverse shock component. In fact, three bursts with an 
identified prompt optical flash, i.e. GRBs 990123 (Akerlof et al. 1999), 
021211 (Fox et al. 2003; Li et al. 2003), and 041219A (Vestrand et al. 
2005; Blake et al. 2005), 
are all modeled to have a low density environment 
(Panaitescu \& Kumar 2002; Kumar \& Panaitescu 2003; Fan et al. 2005). 

It could be that the prompt optical radiation of reverse shocks is 
suppressed relative to that in the simple hydrodynamical models if 
the ejecta is Poynting dominated (e.g. Usov 1992; Thompson 1994; 
M\'esz\'aros \& Rees 1997a, 1997b; Spruit et al. 2001; Lyutikov \& 
Blandford 2003),
since in that case the sound speed, or Alfv\'en speed in the ejecta is
close to the speed of light and the reverse compression wave fails to
steepen into a shock. If in some bursts the outflows were indeed Poynting
dominated, a significant suppression of the reverse shock emission would
be expected (Zhang \& Kobayashi 2005). Conventional analyses of reverse 
shocks have in fact indicated that the ejecta may be more magnetized than 
the forward shock region (Zhang, Kobayashi, \& M\'esz\'aros 2003; Fan 
et al. 2002; Kumar \& Panaitescu 2003), but in general, there is no strong 
observational support for the presence of Poynting dominated flows. 
However, more detailed investigations of purely hydrodynamical reverse 
shocks (i.e. not Poynting dominated) indicate that the strength of the 
optical emission from reverse shocks may be weaker than previously calculated 
(e.g. Kobayashi 2000; Nakar \& Piran 2004; Beloborodov 2005; Kobayashi et al. 
2005; 
Uhm \& Beloborodov 
2006; MacMahon et al. 2006). Thus, there are plausible physical reasons for 
which in the seven GRBs with tight very early optical limits, the reverse shock 
may be considerably suppressed.

Figure 3 highlights three Swift bursts, GRBs 050223, 050421, and 050422 
which are undetected optically and do not fall within the gray band. These 
bursts are deficient in X-ray emission compared to their $\gamma$-ray 
fluxes, as well as being optically faint. GRBs 050421 and 050422, in 
particular, were observed very early ($<$ 95s) after the burst, and were 
not detected by the UVOT to limits roughly 4-5 magnitudes fainter than was 
previously possible (the observational evidence points strongly towards
galactic dust absorption suppressing the optical afterglow in these two bursts). 
These GRBs appear to have a higher than normal $\gamma$-ray efficiency. 
In the case of GRB 050223 the $\gamma$-ray efficiency is estimated to be 
greater than 90\% following the method derived by Lloyd-Ronning \& Zhang (2004). 
A concern, related to the
$\gamma$-ray efficieny of bursts, is the uncertainty in the constant nature 
of the kinetic energy of some bursts. Recent work suggests that the X-ray 
shallow decay phase is caused by energy injection (Rees \& Meszaros 1998; 
Z06a; N06; Panaitescu et al. 2006a; Panaitescu et al. 2006b). Although this 
phase is typically over by 
1 hour, there are cases in which the shallow decay phase continues for longer 
(i.e. GRB 050401; N06). By taking the afterglow kinetic energy 
after injection is over, one gets a moderate $\gamma$-ray efficiency for most of 
the bursts (Z06b). As a result, if in some bursts the energy injection phase 
lasts much longer than 1 hour, taking the kinetic energy derived from 1 hour 
would considerably underestimate the total kinetic energy of the burst. This 
would potentially inflate the $\gamma$-ray efficiency (Z06b). This statement is 
only valid if one believes that the total afterglow kinetic energy after the 
injection is over should be used to define the $\gamma$-ray efficiency. On the 
other hand, it could be that the kinetic energy right after the burst is the 
relevant quantity to define the efficiency. If this is the case, most of the 
$\gamma$-ray efficiencies derived at 1 hour are under-estimated. Nonetheless, 
those with an extended injection phase would have measured efficiencies higher 
than the others, which can account for the apparent high-efficiency bursts in 
our sample. For a more detailed discussion of GRB efficiency, we refer the 
readers to Z06b.

\section{Conclusion}
We have compared the X-ray afterglow flux to the optical afterglow
flux and prompt $\gamma$-ray fluence of a sample of Swift GRBs to 
quantify the lack of optical afterglow emission. Using this 
quantificaton method, a few of these bursts manifest an apparent 
higher than normal $\gamma$-ray efficiency. Although most bursts
proceed through the normal decay phase after 1 hour, some bursts 
manifest energy injection after this time thus varying the kinetic 
energy of the fireball and potentially inflating the $\gamma$-ray 
efficiency. Further work is required in order to determine the true 
$\gamma$-ray efficieny of these bursts. Among various
interpretations for the tight early UVOT upper limits
(e.g. extinction, high redshift, etc), most would not lead to the
observed high $\gamma$-ray efficiency, since they only suppress the
optical emission but not the X-ray emission. 

An investigation into the possible mechanisms for the lack of optical 
emission points to $\sim 25\%$ of the bursts in our sample being 
extincted by galactic dust and $\sim 25\%$ are obscurred by circumburst 
absorption (which is consistent with the $\sim 20\%$ proposed by 
Jakobsson et al. 2006). An additional $\sim 30\%$ are most likely attributable
to Ly-alpha blanketing and absorption at high redshift, although other
mechanisms can not be conclusively ruled out at this time. Rapid temporal 
decay as a valid mechanism is ruled out assuming that the decay profile of 
these optically deficient bursts behave similarly to their optical counterparts.
A low density environment is also eliminated as a possibility based on
comparisons with optically detected afterglows generated in a low 
density environment. While Poynting-flux dominated outflows could
suppress the early-optical afterglow, there are also indications that 
in normal hydrodynamical outflows the physics of the reverse shocks can,
in many cases, lead to weaker optical emission than previously thought.
More data and more detailed modeling is needed to test such explanations
for the paucity of detected optical flashes. 


\acknowledgments
We appreciate the comments from an anonymous referee that greatly
improved the conclusions of this paper. We gratefully acknowledge the 
contributions from members of the Swift team at PSU, MSSL, University 
of Leicester, NASA/GSFC, INAF-OABr, and our subcontractors, who helped make this 
Observatory possible. We are also grateful to the Flight Operations 
Team for their support above and beyond the call of duty. This work is 
supported at Penn State by NASA contract NAS5-00136, at MSSL and 
Leicester by funding from the Particle Physics and Astronomy Research 
Council (PPARC), and at INAF-OABr by ASI contract I/R/039/04. The Los 
Alamos
National Laboratory is operated by the University of California for
the US Department of Energy (DOE).

\clearpage
\begin{table*}
\scriptsize
\caption{Basic properties of the 19 bursts in the Swift UVOT sample}
\begin{tabular}{llllllllllll}
\hline\hline
GRB\tablenotemark{*}&$\log{F_{x,1}}$&$\log{S_{\gamma}}$&$\log {f_{R,1}}$&$t_{U}$&
$\Delta T_{U}$& $m_{U}$&$A_V$&$\Delta T_{G}$&$m_G$&$z$&Refs.\\
&&&($\mu {\rm Jy}$)&(s)&(hrs)&(mag)&(mag)&(days)&(mag)&&\\
(1)&(2)&(3)&(4)&(5)&(6)&(7)&(8)&(9)&(10)&(11)&(12)\\
 \hline
{\bf 050215B}&1.70&-0.64&1.76&100&0.485&18.44&0.08&0.41&K=20.2&-&a,b,c\\

{\bf 050219A}&1.48&0.97&0.63&100&0.026&18.10&0.50&-&-&-&d,e\\

{\bf 050219B}&2.58&1.36&2.27&100&0.871&17.81&0.09&4.14&K$_S$=21.5&-&f,g,h,i,j\\

{\bf 050223}&0.13&-0.13&2.11&100&0.773&18.08&0.28&-&-&-&k,l\\

{\bf 050315}&2.42&0.63&0.66&100&0.024&17.94&0.15&0.48&R=20.9&1.95&m,n,o,p\\

{\bf 050326}&2.28&1.28&2.01&100&0.905&18.51&0.12&-&-&-&q,r,s\\

{\bf 050410}&1.53&0.84&2.10&100&0.528&17.70&0.34&-&-&-&t,u\\

{\bf 050412}&1.34&0.32&0.57&100&0.026&18.25&0.06&-&-&-&v,w\\

{\bf 050421}&-0.71&-0.74&1.57&100&0.031&15.95&2.51&-&-&-&x\\

{\bf 050422}&0.33&0.07&2.50&100&0.031&13.62&4.31&-&-&-&y,z,aa\\

{\bf 050502B}&1.17&-0.11&0.49&100&0.017&17.98&0.09&0.02&I=19.8&-&ab,ac,ad,ae\\

{\bf 050505}&2.79&0.61&2.22&100&0.782&17.81&0.06&0.32&R=21.5&4.27&af,ag,ah,ai,aj\\

{\bf 050509A}&1.92&-0.34&1.23&91&0.014&15.95&1.86&-&-&-&ak,al\\
\hline

050318&2.25&0.32&2.08&100&0.910&17.25&0.06&0.34&R=\#&2.80&am,an,ao,ap\\
050319&2.58&-0.10&1.81&100&0.024&17.09&0.03&0.33&B=20.9&3.24&aq,ar,as,at,au,av\\

050406&0.54&-1.02&0.96&100&0.024&19.41&0.06&0.32&r$\sim$22&2.44&aw,ax,ay,az\\

050416A&2.52&-0.42&1.44&100&0.022&18.90&0.09&0.64&R=20.9&0.65&ba,bb,bc,bd\\

050525A&2.26&1.30&2.65&100&0.024&13.19&0.31&0.22&B=18.8&0.69&be,bf,bg,bh\\

050603&2.93&1.11&4.27&100&9.230&18.01&0.09&0.14&R=16.5&2.82&bi,bj,bk,bl,bm\\
\hline
\hline
\end{tabular}

\tablenotetext{*}{Those marked by bold face are GRBs with no UVOT
counterparts identified.}

\tablecomments{Col. (1): GRB name; col. (2): X-ray flux in the 2-10 keV range at one
hour after the BAT trigger (hereafter referred to as the trigger), in
units of $10^{-13} {\rm erg} \ {\rm cm}^{-2} {\rm s}^{-1}$; col. (3):
prompt $\gamma$-ray fluence in the 15-350 keV range, in units of
$10^{-6} {\rm erg} \ {\rm cm}^{-2}$; col. (4): upper limit on the
UVOT flux extrapolated to the R-band at one hour after the trigger when 
no counterparts are identified and the UVOT flux extrapolated to the 
R-band at one hour after the trigger when a counterpart was found; 
col. (5): exposure time of the UVOT finding chart; col. (6):
time after the trigger for beginning of UVOT finding chart; col. (7):
R-band upper limit when no counterparts are identified and the R-band 
magnitude when a counterpart was identified; col. (8): the amount of 
galactic absorption in the line of sight; col. (9): time after
the trigger before a ground based observation detected the afterglow
at; col. (10): the magnitude in the listed filters (the filters listed 
are the bluest detections as found in the GCN Circulars; \# - Although a 
detection was claimed, no magnitude was provided in the GCN Circulars 
for GRB 050318); col. (11): redshift
determined from ground-based or UVOT data; col. (12): references: (a)
Page et al. 2005; (b) Roming et al. 2005b; (c) Tanvir et al. 2005a;
(d) Romano et al. 2005; (e) Schady et al. 2005a; (f) Burrows et
al. 2005b; (g) Cummings et al. 2005a; (h) Ivanushkina et al. 2005; (i)
D'Avanzo et al. 2005; (j) Berger, Kelson, \& Gonzalez 2005; (k) Giommi
et al. 2005; (l) Gronwall et al. 2005; (m) Parsons et al. 2005a; (n)
Rosen et al. 2005a; (o) Cobb \& Bailyn 2005a; (p) Kelson \& Berger
2005; (q) Markwardt et al. 2005; (r) Cummings et al. 2005b; (s)
Holland et al. 2005; (t) La Parola et al. 2005; (u) Boyd et al. 2005a;
(v) Cummings et al. 2005c; (w) Roming et al. 2005c; (x) Barbier et
al. 2005a; (y) Barbier et al. 2005b; (z) Suzuki et al. 2005; (aa)
McGowan et al. 2005a; (ab) Falcone et al. 2005; (ac) Cummings et
al. 2005d; (ad) Cenko et al. 2005a; (ae) Cenko et al. 2005b; (af)
Hurkett et al. 2005a; (ag) Hullinger et al. 2005; (ah) Rosen et
al. 2005b; (ai) Tanvir et al. 2005b; (aj) Berger et al. 2005; (ak)
Hurkett et al. 2005b; (al) Barbier et al. 2005c; (am) Krimm et
al. 2005a; (an) McGowan et al. 2005b; (ao) Mulchaey \& Berger 2005;
(ap) Still et al. 2005; (aq) Rykoff, Schaefer, \& Quimby 2005; (ar)
Krimm et al. 2005b; (as) Boyd et al. 2005b; (at) Sharapov et al. 2005;
(au) Fynbo et al. 2005; (av) Mason et al. 2006; (aw) Parsons et
al. 2005b; (ax) Rol et al. 2005b; (ay) Berger, Oemler, \& Gladders
2005; (az) Schady et al. 2006b; (ba) Sakamoto et al. 2005; (bb) Schady
et al. 2005b; (bc) Kahharov et al. 2005; (bd) Cenko et al. 2005c; (be)
Rykoff, Yost, \& Swan 2005; (bf) Cummings et al. 2005e; (bg) Cobb \&
Bailyn 2005b; (bh) Blustin et al. 2005; (bi) Retter et al. 2005; (bj)
Fenimore et al. 2005; (bk) Brown et al. 2005; (bl) Berger \& McWilliam
2005; (bm) Berger \& Becker 2005.}

\end{table*}

\clearpage

\begin{figure}
\plotone{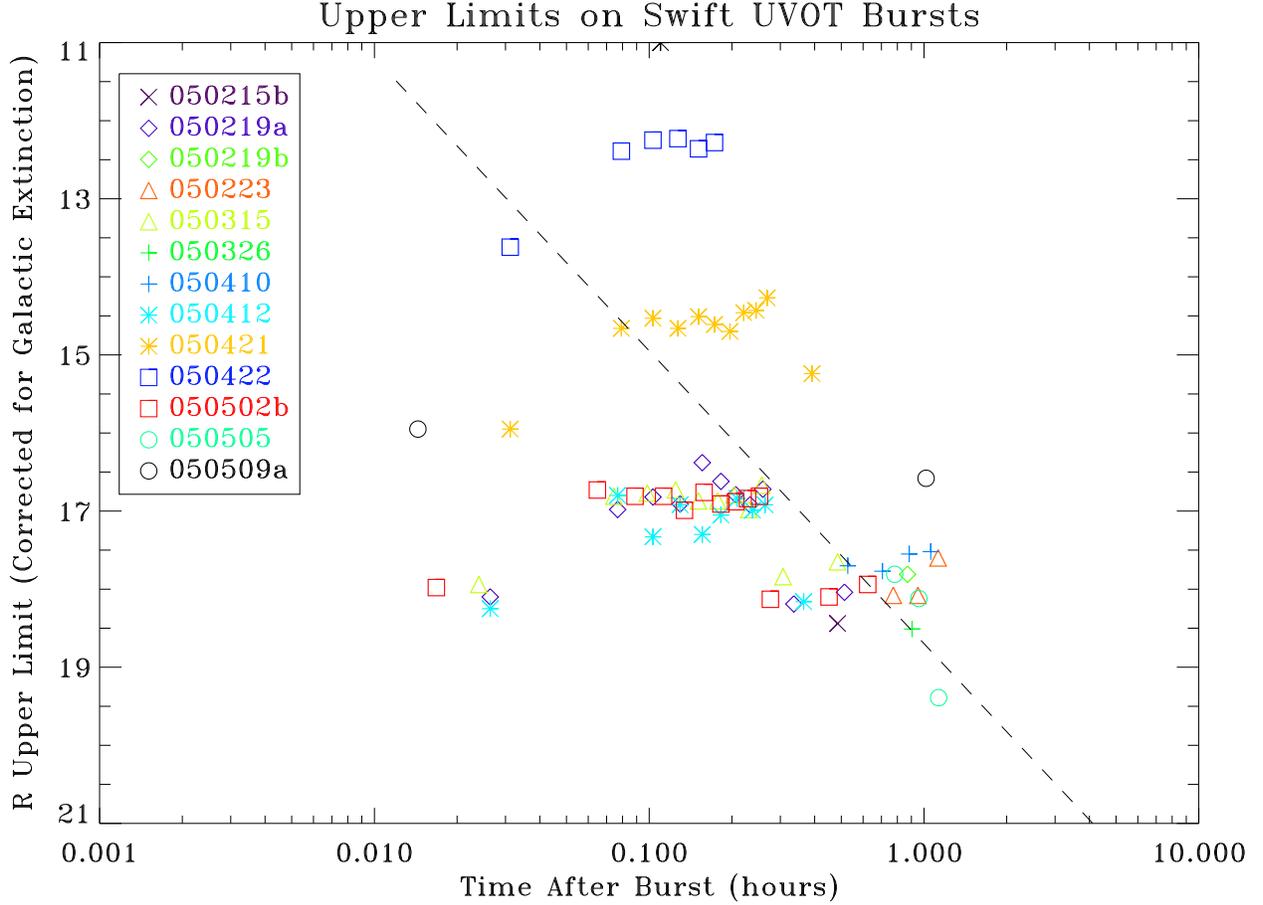} 
\caption{The 3$\sigma$ UVOT-magnitude upper limit for each exposure
corrected for galactic extinction and adjusted to the R-band. The UVOT
prompt slew observation sequence consists of a settling exposure of
$\sim$10 s in the UVW2 or V filter, a finding chart exposure of 100 s
in the V filter, and 10, 100, and $\sim$900 s exposures in all the
broadband filters for a total exposure time of $\sim$0.6, $\sim$3, and
$\sim$22 ks, respectively. The dotted line represents the dividing
line used by Rol et al.(2005a), extrapolated to earlier times, that
separates the dark bursts in their sample from practically all
detected pre-Swift afterglows.\label{fig1}}
\end{figure}

\begin{figure}
\includegraphics[angle=-90,scale=0.7]{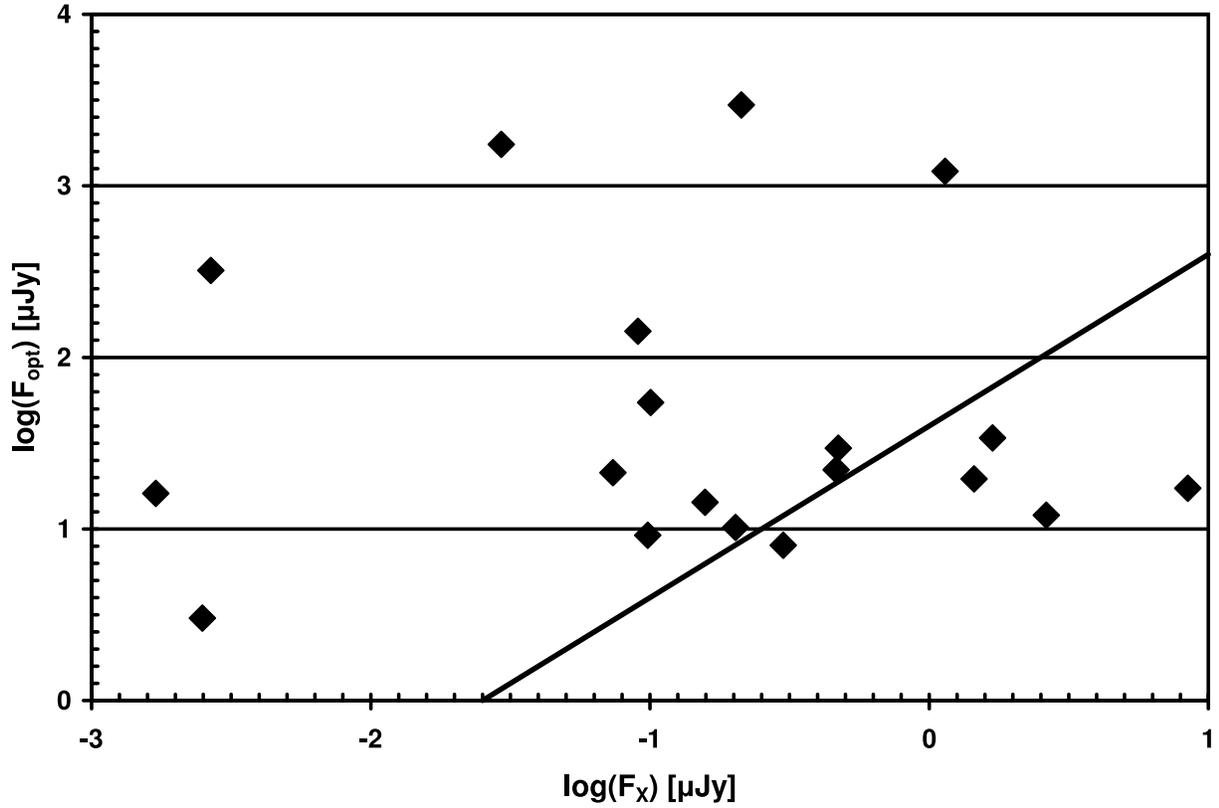}
\caption{The optical-to-X-ray spectral index
($\beta_{OX}$) for the 19 bursts in the Swift UVOT sample. The
dividing line represents $\beta_{OX}=0.50$ which separates the dark
bursts from the non-dark bursts, as defined by Jakobsson et al. (2004). 
The bursts below the dividing line
are GRBs 050315, 050319, 050326, 050412, and 050505.\label{fig2}}
\end{figure}

\begin{figure}
\epsscale{.9}
\plotone{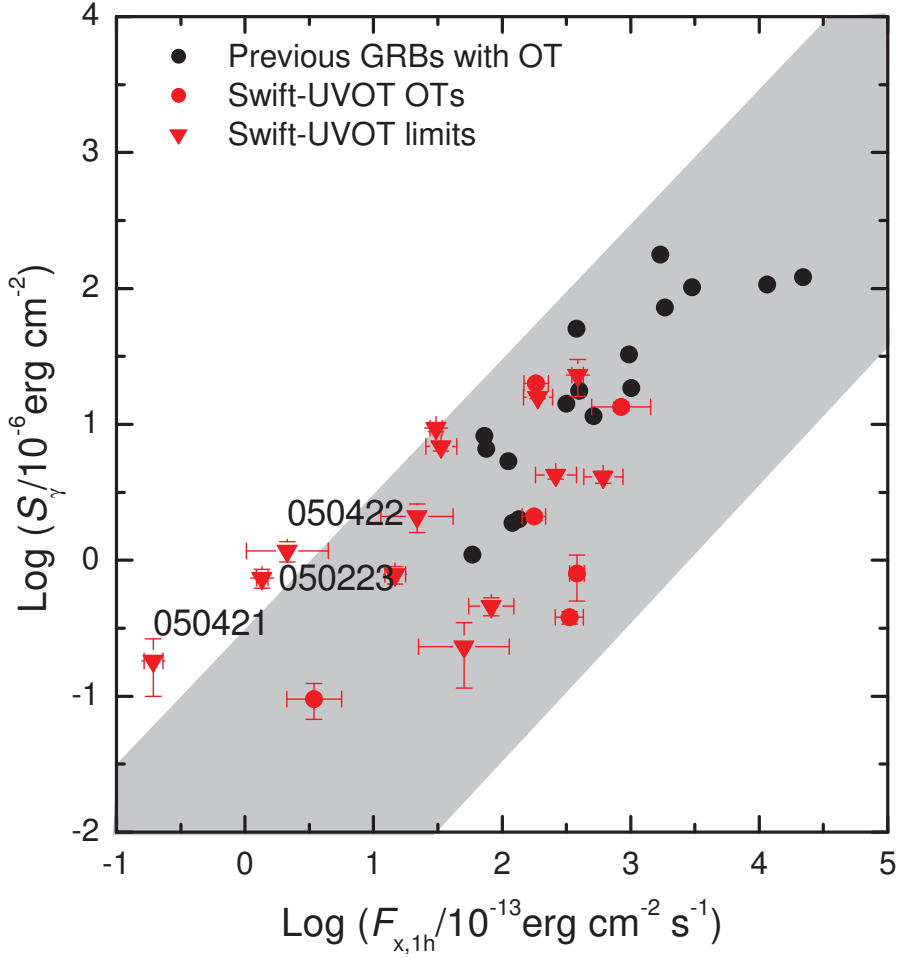} 
\caption{Distribution of Swift GRBs in the
$F_{x,1}-S_{\gamma}$ plane.
$F_{x,1}$ is 
the X-ray afterglow flux in the 2-10 keV band 
at 1 hour after the burst.
$S_{\gamma}$ is the
$\gamma$-ray fluence in the 15-350 keV energy band using the best-fit
spectral parameters for Swift GRBs, or assuming a break energy at 250
keV with the low and high photon energy indices set to -1 and -2.3,
respectively, for GRBs observed with other telescopes. 
Red and black circles represent the Swift GRBs and GRBs observed with
other telescopes, respectively. 
Solid triangles denote those Swift GRBs undetected by UVOT. For
non-Swift bursts, the X-ray data 
are taken from the literature (Berger, Kulkarni, \& Frail 2003). 
The X-ray value for GRB 050421 was 
calculated by extrapolating from the 100 to 500 second interval to 1 
hour. The extrapolated value is consistent with observational upper 
limits on either side of 1 hour. The shaded region for
the panel is twice the standard deviation ($2\sigma$) of
$S_{\gamma}/F_{x,1}$ 
for optically-detected GRBs.\label{fig3_01}}
\end{figure}

\begin{figure}
\plotone{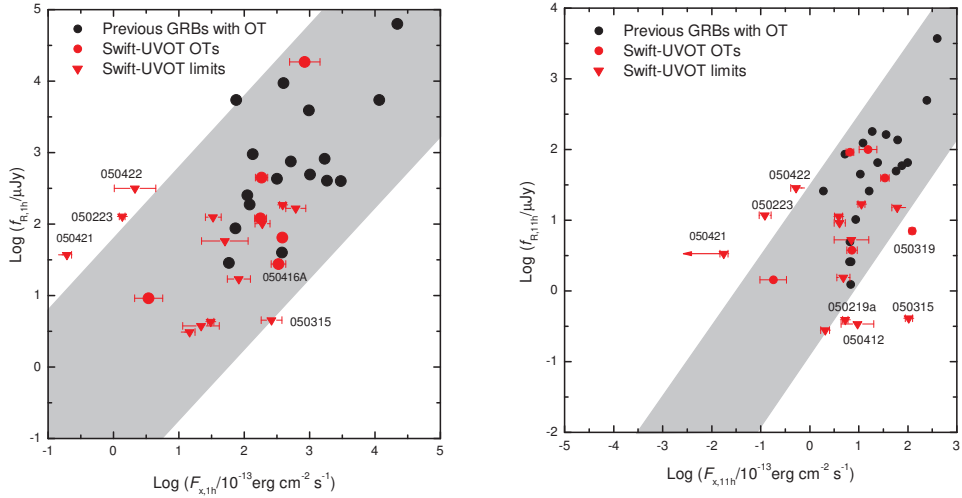} 
\caption{Distribution of Swift GRBs in the
$F_{x}-f_{R}$ plane at 1hr (left panel) and 11hrs (right panel) after
the burst trigger, where $f_{R}$ is the R-band flux.
Symbols are similar to those in Figure 3. The 1 hr optical upper limits 
for bursts without UVOT observations are derived by assuming a temporal 
decay index of -1 and the upper limits obtained by the ground-based 
telescopes. For three Swift bursts, GRBs 041223, 050117, and 050126, 
the upper limits are taken from Covino et al. (2005), Karimov et al. 
(2005), and Lipunov et al. (2005), respectively. For the 
pre-Swift bursts, these limits are taken from 
J04. The X-ray value at 11 hours for GRB 050421 was 
calculated by extrapolating from 1 hour using a slope of -1.
The shaded region
for the panels is twice the standard deviation ($2\sigma$) of
$F_{x}/f_{R}$ for optically-detected GRBs.\label{fig3_02}}
\end{figure}

\begin{figure}
\plotone{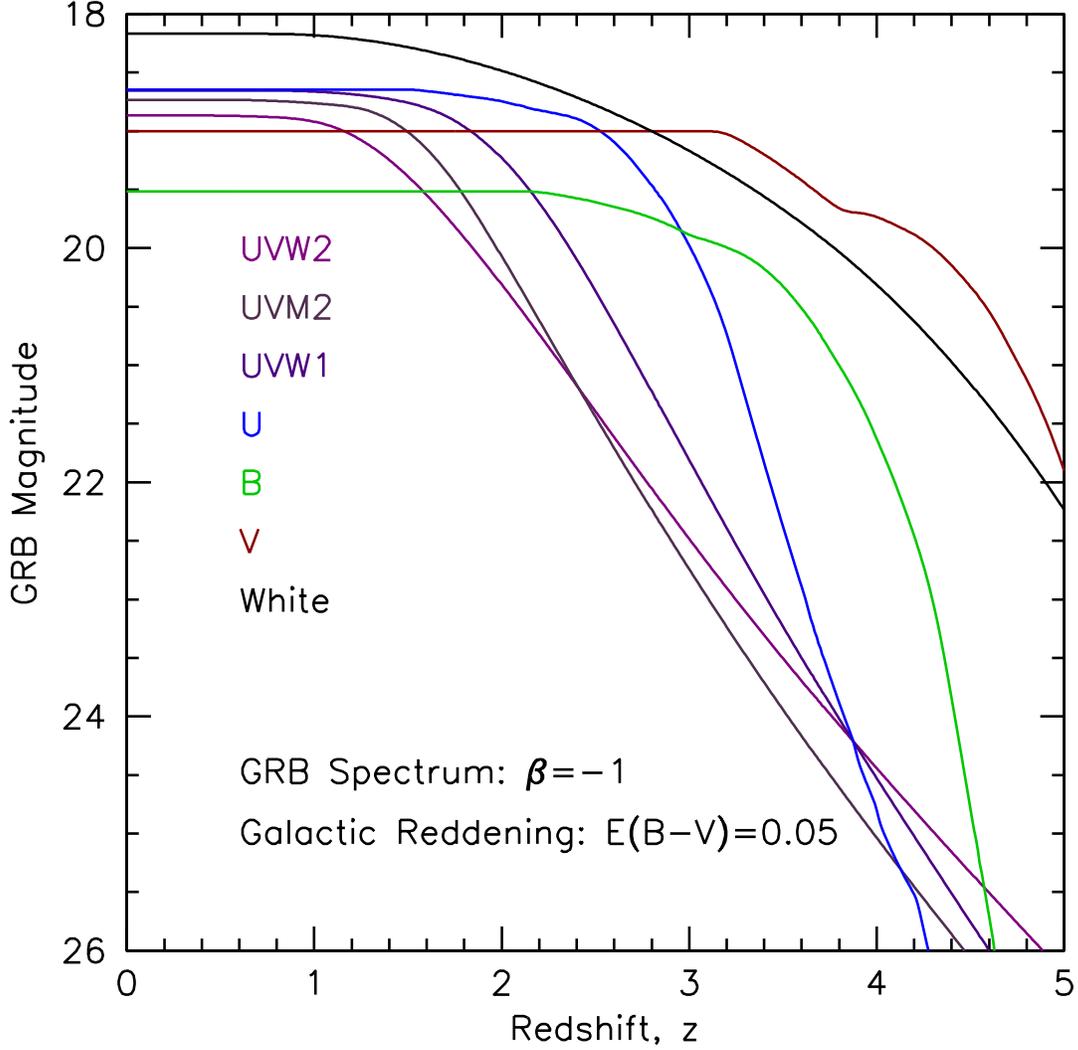} 
\caption{UVOT magnitude of a GRB vs. redshift for a ``typical GRB" 
spectrum. A power-law SED, with spectral index $\beta = -1$ and Galactic 
$E(B-V) = 0.05$, that is normalized to have $V = 19.0$ at $z = 0$ was
assumed. The current UVOT filter effective area curves, zero-points, and 
counts-to-flux density calibrations were used derive the magnitudes. In 
the absence of Lyman absorption, the magnitudes would remain as they are 
at $z = 0$. The spectrum is then run through Lyman extinction as a 
function of redshift, according to Madau (1995). The initial ($z = 0$) 
colors are due to $\beta$ and $E(B-V)$, but the falling magnitudes are 
due to Lyman absorption (Ly-$\alpha$, Ly-$\beta$, Ly-$\gamma$, Ly-$\delta$, 
Ly-continuum).\label{fig5}}
\end{figure}

\end{document}